\documentclass[12pt]{article}
\usepackage{amsmath}
\usepackage{amsfonts}
\begin{document}
\def\eg{{\em e.g.\,}}
\def\ie{{\em i.e.\,}}
\def\C{\mathrm{C}}
\def\Q{\mathrm{Q}}
\def\cH{\mathcal{H}}
\def\dof{{d.o.f.\,}}
\title{What can biology bestow to quantum mechanics?}
\author{M.V.Altaisky\thanks{Joint Institute for Nuclear Research,
 Dubna, 141980, Russia;
 and Space Research Institute RAS, Profsoyuznaya 84/32, 
Moscow, 117810, Russia, altaisky@mx.iki.rssi.ru.}}
\date{}
\maketitle
\begin{abstract}
The biological hierarchy and the differences between living and non-living 
matter are considered from the standpoint of quantum mechanics.
\end{abstract}
\section{Quantum Mechanics and Evolution}
The most important discovering in natural sciences are in some or 
other way connected to quantum mechanics. There is also a bias that 
biological phenomena will be explained by quantum theory in future, 
since quantum theory already contains all basic principles of particle 
interactions and these principle had success in molecular dynamics, the 
basis of life. Nevertheless, it seems there is one concept in biology 
that could be hardly found in quantum physics. This is the concept of 
{\sf Evolution}.

Evolution is often identified with {\sf self-organization}. At least they 
have much in common. The  self-organization is observed in both living 
systems, and non-living ones, \eg in Belousov-Zhabotinsky reaction. 
It is noteworthy, that the {\em approximation of the dissipative system} 
simply means that the system is open, having income and outcome energy flux,
 and some of its degrees of freedom 
(\dof) are not considered explicitly, instead only the energy flux going 
out of the system (say, in the form of radiation) via that \dof is taken 
into account. 

Very often the biological studies are regarded as opposite to the physical 
ones in the sense that they are qualitative rather than quantitative. At the 
same time the biology yields a number of concepts and basic facts those 
are not displayed explicitly in inanimate phenomena. We suggest 
the following three facts to be of principle importance:
\begin{enumerate}
\item   The properties of a living system are more than just a collection 
        of its component properties. In other words, it is impossible to 
        predict the whole set of properties of a complex biological system 
        even having known all properties of its components and interactions.
\item   The properties and functions of the components of a system depend 
        on the state of the whole system. In other words, the same components 
        being included in different systems may have different properties.
\item   There is an {\sl Evolution} --- a process of creating new entities,
        forms and functions on the base of the existing components. 
\end{enumerate} 
Thus, at least one thing is common for biology and quantum physics: 
in both the relation {\sf ``the part - the whole''} is extremely important. 
And in both this relation is not trivial. 
$$ \Psi(x) = \prod_i \psi_i(x_i) $$ only for the systems of 
noninteracting particles. 

The other thing is not so important. The properties (2,3) are implicitly 
based on the {\sf concept of scale}: an entity to become a part of another 
entity should be in some metric (not necessary Euclidean) smaller than it. 
If the metric is Euclidean, or at least Archimedian, the evolution of the 
Universe can be said to go from small scales to large scales, in accordance 
to the Big Bang scenario. In this sense, the elementary particles and atoms 
do have (or had had) their evolution: at early times of the Universe  the 
nucleons had been built of quarks, the nuclei from nucleons and so long. 
The same is true for not so far geological history: the minerals and crystals 
evolved from atoms and molecules. 

We do not have an ultimate answer for the question, 
why the evolution had taken the way it has been going through.
However, {\em if the whole is more than the sum of parts and the properties 
of the parts depend on the state of the whole, there are some implications 
for quantum mechanics}.

To describe a state (in common sense of quantum mechanics) of an object 
$A_1$ (interacting with objects $A_2,\ldots,A_N$) which is a part of an object 
$B_1$ we have to write the wave function in the form 
\begin{equation}
\{ \Psi_{B_1},  \Psi_{B_1A_1} \},
\label{eq1}
\end{equation}
where $\Psi_{B_1}$ is the wave function of the whole (labeled by $B_1$), 
and $\Psi_{B_1A_1}$ is the wave function of a {\sl 
component} $A_1$ belonging to the entity $B_1$.
For instance, $A_1,A_2,A_3$ may be quarks, and $B_1$ may be proton.
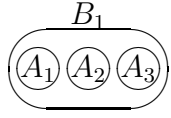
\begin{figure}
\begin{center}
\begin{picture}(80,50)
\put(40,40){\makebox(0,0){$B_1$}}
\put(40,20){\oval(60,30)}
\put(21,20){\circle{17}}
\put(40,20){\circle{17}}
\put(59,20){\circle{17}}
\put(21,20){\makebox(0,0){$A_1$}}
\put(40,20){\makebox(0,0){$A_2$}}
\put(59,20){\makebox(0,0){$A_3$}}
\end{picture}
\end{center}
\caption{The whole $B_1$ and its parts $A_1,A_2,A_3$.}
\end{figure} 
The objects $A_1,\ldots,A_N$ are {\em inside} $B_1$ and hence it is 
impossible to commute $[\Psi_{B_1},\Psi_{b_1A_1}]$ or 
to multiply them $\Psi_{B_1},\Psi_{B_1A_1}$. {\em The functions 
$\Psi_{B_1}(x)$ and  $\Psi_{B_1A_1}(x)$, taken in coordinate representation, 
live in different functional spaces}. To label the hierarchical object 
(\ie to set a coordinates on it) one needs a hierarchical tree, like those 
used in biology to trace the evolution. 

What we generally observe, is 
that {\sf each hierarchical level has its own symmetry group}. This is $SU_3$ 
for quark level or isospin group for nuclei. So, each hierarchy level 
should be described by triplet 
$$I,G_I,X^{G_I},$$ 
where $I$ is merely a label for the scale, $G_I$ is the symmetry group at 
this scale, $X^{G_I}$ is a topology on $G_I$, or coordinates on the $I$-th level.  
The wave function of an object $B^{I_1}$ of the level $I_1$ consisting 
of $N$ objects $\{A_i^{I_2}\}_{i=1,N}$ can be written as 
\begin{equation}
\Psi_B =  \Bigl\{ 
        \psi^{I_1}_B(x^{G_{I_1}}),
        \{\psi^{I_2}_{BA_1}(x^{G_{I_2}}_1),\ldots,
        \psi^{I_2}_{BA_N}(x^{G_{I_2}}_N) \},\ldots
        \Bigr\}.
\label{hier}
\end{equation}
The Euclidean space is a particular case of the translation group 
$$G_I : x \to x +b.$$

In both, physics and biology, the symmetry breaking plays an 
important role. It is known, that the amount of information written 
in DNA, if calculated as one nucleotide -- one bit, is far from being 
sufficient to describe  the formation of adult organism. Therefore, 
the information is likely to be written more effectively than just 
a technical plan of the organism. What is encoded, is probably a chain 
of bifurcation points to be undergone in growth process. What is observed is 
a {\sf hierarchy of symmetry breaking}, which can be described 
as a change of topology. 
If the quantum mechanics is valid on the macroscopic, \ie on the organism 
level, we can say that the higher level of the hierarchy emerge as a result 
of

a) self-organization \cite{prigogine};

b) auto-evolution \cite{ldf};

c) $\{ \emptyset; \psi_{A_1} \otimes \psi_{A_2}\} \to 
   \{ \psi_{B}; \{ \psi_{BA_1},\psi_{BA_2}\}\}$.

By empty-set $\emptyset$ at the left hand side of the latter 
equation we denote the non-existing common ``container'' for two 
components $\psi_{A_1}$ and $\psi_{A_2}$; $\psi_{B}$ is a new entity 
formed by $\psi_{A_1}$ and $\psi_{A_2}$. By no means we are saying that the 
object $\psi_{B}$ exists {\em before} its components, we just say that the 
wave function $\psi_{B}$ should exist as a {\sf possibility} for its potential 
components to join each other. In this sense the possibility emerge first. 

The hierarchy of biological system joints the hierarchy of non-living matter 
by means of {\sl genom} -- the sequence of macromolecules which prescribes 
the evolution of all living systems, from cell to organism.
\vskip0.1cm
\begin{center}
{\sf The hierarchy levels of living and non-living matter}
\vskip0.1cm
\begin{tabular}{|r|r|}
\hline
Living matter & Non-living matter \\
\hline 
\dots     & \\
ecosystem &            \\
population&  \\
organism  & \\
organ     & \\
cell      & \\
organell  & \\
\multicolumn{2}{|c|}{genome} \\
          & molecule \\
          & atom \\
          & nucleus \\
          & nuclon \\
          & \dots  \\
\hline
\end{tabular}
\end{center}  
The place on an entity on the hierarchy tree and its distance from the 
position of the organism it belongs to determines the dynamical 
repertoire of the entity. The evolutional distance between maximal and 
minimal parts of the organism determines its ability of self-recovering. 

If one end of {\em Hydra oligactis} (a simplest animal living in water) 
is cut off, the remaining cells react to the absence of this part by 
rearranging themselves, giving growth to new cells and form a complete 
animal. This process involves at least three levels: \\
\centerline{Organism $\longrightarrow$ Cell $\longrightarrow$ Cell component.}    
The above described process of self-repairing can be described 
by following diagram
\begin{eqnarray}
\{ \psi_A; \{\psi_{AC_1}, \ldots , \psi_{AC_N} \}\} + \Gamma \to
\{ \psi_A; \{\psi_{AC_1}, \ldots , \psi_{AC_K} \}\} +
\{\psi_{C_{K+1}}, \ldots , \psi_{C_N} \} \label{diss}\\ 
\to \{ \psi_A; \{\psi_{AC_1c_1},\psi_{AC_1c_2} \ldots , \psi_{AC_Kc_L} \}\}
\label{restr} \\ 
\to \{ \psi_A; \{\psi_{AC_1}, \ldots , \psi_{AC_N} \}\} 
\label{recover}
\end{eqnarray} 
On the first stage (\ref{diss}) affected by destructive action $\Gamma$, 
the part of the system, a block of $(N-K)$ cells is cutted of. The 
{\sl remainder} 
\begin{equation}
\{ \psi_A; \{\psi_{AC_1}, \ldots , \psi_{AC_K} \}\}
\label{remainder}
\end{equation}
 does not form a complete organism any longer; the product of representations 
$\displaystyle \bigotimes_{i=1}^K T(G_{C_i})$ 
does not contain  $T(G_{A})$. 
So, the wave function of the remainder (\ref{remainder}) breaks down 
to the third level hierarchy wave function  
$$\{ \psi_A; \{\psi_{AC_1c_1},\psi_{AC_1c_2} \ldots , \psi_{AC_Kc_L} \}\},$$
which provides a possibility of building a  representation tensor product, 
which contains $T(G_{A})$. On this third level the wave functions are 
being rearanged 
according to this tensor product and the missed second level blocks rebuild.

The process (\ref{diss}--\ref{recover}) clearly has its physical 
counterpart in the recombination taking place after photoionization 
in gaseous media. 
\begin{eqnarray}
A+\gamma &\to& A^{+} + e_{(1)}\label{d} \\
A^{+} + e_{(2)} &\to& A \label{r}
\end{eqnarray}
The stage (\ref{d}) is the ionization of the atom $A$, 
the stage (\ref{r}) is the recombination by capturing an electron $e_{(2)}$ 
from the {\em environment}. On the first stage the symmetry group $G$ of 
the atom $A$ breaks down to that of the ion $A^+$ and electron $e_{(1)}$
$$G_A \to G_{A^+}\otimes (SU_2 \otimes U_1)_e,$$
at the second stage the symmetry is restored. 

The observations at all levels of the evolutional hierarchy, from simple 
organnels to complex ecosystems show that {\sf only neighboring 
levels can interact}. The quantum nature of the interactions 
in this hierarchy could be used to understand, why a cutted skin recovers, but 
the arm cutted of is not being recovered. If the lost part of the organism 
has $M\gg1$ hierarchical levels a $M$-level cascade process should run 
in the remainder to rebuild the lost part. A process of this type will 
require ({\em a}) a significant flux of energy, and at the same time  
a tremendous flux of negative entropy, required to restore the symmetry
of the wave function of the whole organism by rearange the wave 
functions of its components. This combination is hardly possible.   

Another fact, that can be considered as an argument for quantum nature 
of the evolution is the {\sf memory of the cell} \cite{cell}. A clone 
of eucaryotic cells growing, some of the cells becomes differentiated 
from each other and acquire different functions. Physically, the 
differentiation goes on in response to the action of neighboring 
cells and external factors. However, it is remarkable that {\em most 
eucaryotic cells usually persist in their specialized states after 
the influences the differentiation was caused by are eliminated}. This 
can be thought of as the cells being in $\psi_{AC}$ state rather than 
in $\psi_C$.
\section{Pauli principle}
Let us recall the Pauli principle. {\em Two electrons of the same atom 
can not be in the same quantum state}. Or more generally: {\em Two 
fermions of the same system can not be in the same quantum state}. 
The latter is a direct consequence of the fact that the wave function of 
a fermion system should be antisymmetric with respect to particle 
transposition; if two of the particles are identical,\ie are in 
the same quantum state, the 
wave function should be both symmetric and antisymmetric, and hence is
exactly zero.

If we suggest the interactions of neighboring levels is important only, 
we can say that, {\sf two fermions belonging to the same system  of the 
next hierarchical level can not be in the same state}. Therefore, it is 
impossible for two electrons in atom to have the same quantum numbers 
($n,l,m$), but is it possible for two electrons 
of the same {\em molecule} to be in the same state?
It seems evident that  
two electrons of different macroscopic objects can be in the same state.
{\em But is it really possible for two electrons of the same molecule?}

So, the real question is: what should we really mean by ``the hierarchy 
level next to atom''?
There is no common sense answer to this question, but if the generalization 
of Pauli principle formulated above, is valid, and the only question is 
what is the {\em next hierarchical level}, the matter can be experimentally 
investigated, at least in principle. To some extent, the idea of 
possible experiment of this type have been already suggested by D.~Home 
and R.~Chattopadhyaya \cite{HC1996}. The setup of their experiment is 
aimed for living systems, so it is discussed in the next section.  

\section{Wave functions of living and non-living systems}
The presence of hierarchical structure of interactions considered 
above in this paper {\em necessary, but not sufficient condition} for 
a system to be alive. For instance, we can imagine a two level 
hierarchical system (see Fig.~\ref{hs:pic} below) with coupling constants of 
the lower level dependent on the state of higher level and {\em vice 
versa}. 
\vskip0.1cm
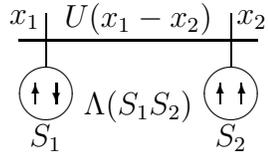
\begin{figure}
\begin{center}
\begin{picture}(100,60)
\put(15,20){\circle{20}}
\put(11,16){\vector(0,1){8}}
\put(19,24){\vector(0,-1){8}}
\put(15,30){\line(0,1){20}}
\put(15,3){\makebox(0,0){$S_1$}}
\put(85,3){\makebox(0,0){$S_2$}}
\put(85,20){\circle{20}}
\put(81,16){\vector(0,1){8}}
\put(89,16){\vector(0,1){8}}
\put(85,30){\line(0,1){20}}
\put(5,40){\line(1,0){90}}
\put(7,48){\makebox(0,0){$x_1$}}
\put(93,48){\makebox(0,0){$x_2$}}
\put(50,48){\makebox(0,0){$U(x_1-x_2)$}}
\put(50,15){\makebox(0,0){$\Lambda(S_1S_2)$}}
\end{picture}
\caption{Toy model of two level hierarchical system}
\label{hs:pic}
\end{center}
\end{figure}

The toy-model Hamiltonians (\ref{hsh}) accounts for the interaction 
of two blocks with coordinates $x_i,i=1,2$, comprised of two 
spins $(s_1^i,s^2_i)$. The blocks are interacting with the potential 
that has position-dependent and spin-dependent parts $W=U+\Lambda$. 
The effective mass of each block depends on the spin-spin interaction 
of its components, with their interaction constant depending on the 
velocity of block as a whole.  
\begin{equation}
\begin{array}{lcl}
H &=& \frac{m_1 \dot{x_1^2}}{2} + \frac{m_2 \dot{x_2^2}}{2} 
   + U(|x_1-x_2|) + \Lambda (S_1 S_2) \\
\\
m_1 &=& m_0 + \lambda(\dot{x_1^2})s_1^1s_2^1 \\ 
m_2 &=& m_0 + \lambda(\dot{x_2^2})s_1^2s_2^2 \\
S_1 &=& s_1^1 + s_2^1 \\
S_2 &=& s_1^2 + s_2^2 
\end{array}
\label{hsh}
\end{equation}

The living and non-living systems are different in the {\sf complexity},
in the Kolmogorov sense, 
of their evolution operators. The Hamiltonian {\sl for a non-living system}
can be constructed using the representations of the symmetry groups of 
its components and their interactions. This description is shorter than 
a time seria of the matrix elements $E_{mn}(t)$ taken at each moment of 
time. {\sl For a living system} the shortest description of the evolution 
operator may be the time seria  $E_{mn}(t)$ itself, or its subseria. 

Let us formulate the difference in the language of group theory:

\begin{enumerate} 
\item We can assert, that {\bf for nonliving systems}  
the knowledge of irreducible representations of the component wave function 
and a symmetry group which accounts for their interaction completely 
determines the wave function of the whole. For example, the wave function 
of nuclon can be obtained since we know it consists of quarks 
($SU(2)$ representation with respect to rotations, 
$SU_F(3)$ internal symmetry - flavor interaction).    
\item In contrast, we assert that the 
wave function of {\bf a living system} is constrained, but not completely 
determined by the representations of symmetry group of its components. 
This means, that even if we know the wave functions of all components 
of a living system we can not predict the behavior of the system without a 
separate knowledge of the next level wave function, i.~e. the wave function of 
the whole. If it is so, a living system should be described by a wave 
function of the type  \eqref{hier}, with $\Psi_B$ being the wave function 
of the whole, $\Psi_{BA_i}$ being the wave functions of the components.
\end{enumerate}
Now, let us turn to the possibility of experimental testimony of these 
two alternatives. If long biological macromolecules, the DNA, can be 
used as a device for quantum measurement \cite{HC1996}, it indirectly 
means that if a photon absorbed by a DNA molecule, the wave function of 
the whole 
molecule flops from one quantum state into another. But the DNA molecule 
itself consists of smaller molecules. So, there are two alternatives: 
either the absorption of a photon changes the wave function of the DNA only 
by changing the wave function of one of its components, or it changes 
the wave function of the whole DNA. If the latter is the case, then due to the 
interaction between the whole and its parts, the absorption of the photon 
at one edge of DNA can can be immediately detected at the opposite edge, 
at least in principle.

To conclude with, we should mention that possible distinction between living 
and non-living systems, itemized above in this paper, makes a new point in 
the Schr\"odinger cat problem 
$$\frac{1}{\sqrt2}|cat\ dead\rangle + \frac{1}{\sqrt2}|cat\ alive\rangle = ?$$
In hierarchical formalism, the wave function of a {\em dead cat} is 
constructed from the direct products of the irreducible representations 
of its parts. The wave function of the {\em alive cat} comprise the 
wave function of the whole cat as well. So these two wave function live in different 
functional spaces. 
  
\vskip\baselineskip
\centerline{***}
The author is thankful to Profs. F.Gareev, T.Gannon and 
Dr. O.Mornev for critical reading of the manuscript and useful comments.
The author also appreciate the financial support and the stimulating 
interest from the Department of Mathematics, University of Alberta, where 
this research was partially performed. 
\appendix
\section*{Appendix}
\subsection*{The structure of the Hilbert space of hierarchical states}
The Hilbert space $\cH$ of hierarchical states \eqref{hier}
may be endowed with a natural structure of vector space:
$$\psi_1,\psi_2 \in \cH,\quad a,b \in \C \Rightarrow 
\psi = a\psi_1 + b \psi_2 \in \cH,$$
where $\psi_1,\psi_2$ are hierarchical wave functions defined by 
\eqref{hier}. 
By definition 
\begin{equation}
a\Psi_B =  \bigl\{ 
        a\psi^{I_1}_B(x^{G_{I_1}}),
        \{a\psi^{I_2}_{BA_1}(x^{G_{I_2}}_1),\ldots,
        a\psi^{I_2}_{BA_N}(x^{G_{I_2}}_N) \},\ldots
        \bigr\}.
\label{hiermult}
\end{equation}
The addition operation  (``{\bf +}'') is defined componentwise:
\begin{equation}
\begin{array}{l}
 \bigl\{
\Phi^1, \{\Phi^2_1,\ldots,\Phi^2_N \},\ldots  
\bigr\}
+ \bigl\{
\Psi^1, \{\Psi^2_1,\ldots,\Psi^2_N \},\ldots  
\bigr\} \\
= \bigl\{
\Phi^1+\Psi^1, \{\Phi^2_1+\Psi^2_1,\ldots,\Phi^2_N+\Psi^2_N \},\ldots  
\bigr\},
\end{array}
\end{equation}
where the group indices of Eq.\ref{hier} are substituted by numbers 
for simplicity.

Nevertheless, it may seem attractive to extend the structure $\cH \to \tilde\cH$ 
by incorporating 
some unphysical states. Let $\{e^1_{i_1}\}$ be the basis for the first  level   
of hierarchy  ($I_1$), let $\{ e^2_{j_1}\otimes \ldots \otimes e^2_{j_N}\}$ be the basis for the 
second level ($I_2$) etc. Then we can define $\tilde\cH$ as linear span  
\begin{equation}
\Bigl\{ \sum_{i_1} a^1_{i_1} e^1_{i_1}, 
\bigl\{ \sum_{j_1}a^2_{j_1} e^2_{j_1}, \ldots ,\sum_{j_N}a^2_{j_N} e^2_{j_N} \bigr\}, \ldots 
\Bigr\}, 
\label{tch}
\end{equation}
where $a^k_j \in \C$ are arbitrary complex numbers. 

Evidently, not all linear combinations of the form \eqref{tch} are physical states. To 
illustrate this, let us consider a system {\sf B} (first hierarchic level) 
constiduted of two subsystems {\sf B1} and {\sf B2} (second hierarchic level). 
Let us ascribe the spin $1/2$ to {\sf B1} and {\sf B2} and write down the possible 
spin states of the whole system. The states 
$$
(1,(1/2,1/2))\quad (-1,(-1/2,-1/2)) \quad (0,(1/2,-1/2)) \quad (0,(-1/2,1/2))
$$
are physically possible. In contrast, the state  
$$ (-1, (1/2,1/2)) $$
is physically impossible, but its structure does not contradict the definition  \eqref{tch}. 

To be more formal, we can use the language of group theory. Let $T^2$ be the irreducible 
representation of $G_{I_2}$ used to construct the basic set  
$\{ e^2_{j_1}\otimes \ldots \otimes e^2_{j_N}\}$. Let the product of representations 
used to construct the second hierarchy level basis be decomposed into direct sum of 
irreducible blocks
\begin{equation}
{\underset{1}{T}}^2 \otimes \ldots \otimes {\underset{N}{T}}^2 = \underset{k}{\oplus} D_k.
\label{irp}
\end{equation}
Then only those basic vectors $e^1_i$ can be considered as {\em physical} (for the first 
hierarchy level, $I_1$), which transform according to one of the irreducible representations 
present in the r.h.s of  
\eqref{irp} (built for the second hierarchy level, $I_2$). Exactly as it takes place in 
the angular momentum theory. 


\begin{thebibliography}{9}
\bibitem {prigogine} Nicolis G., Prigogine I. {\em Self-Organization 
        in Non-Equilibrium Systems}, Wiley, New York, 1977.
\bibitem {ldf} Lima-de-Faria A. {\em Evolution without selection}. 
        Elsevier, 1988.
\bibitem {cell} Alberts B., Bray D., Lewis J., Raff M., Roberts K. 
                and Watson J.D. {\em The molecular biology of the cell}, 
                Garlance Publishing Inc., New York, 1994.
\bibitem {HC1996} Home D. and Chattopadhyaya, R. DNA molecular cousin of 
                  Schr\"odinger's cat: A curious example of quantum 
                  measurement. {\em Phys. Rev. Lett.}{\bf 76}(1996)2837-2839.  
\bibitem {alt} Altaiski, M.V. On some geometrical consequences of 
        the incorporation of Resolution in the Definition of a Coordinate 
        system, {\em Differential Equations and Dynamical Systems},
        {\bf 4}(1996)267-274. \\
        Altaisky M.V., Bednyakov V.A. and Kovalenko S.G. On fractal 
        structure of quantum gravity and relic radiation anisotropy. 
        {\em Int. J. Theor. Phys.} {\bf 35}(1996)253-261.
\end{thebibliography}
\end{document}